\documentclass[prl,floatfix,twocolumn,superscriptaddress,showpacs,preprintnumbers,longbibliography,nofootinbib]{revtex4-1}
\usepackage{dcolumn}    
\usepackage{bm} 
\usepackage{graphicx}
\usepackage{amsmath}    
\usepackage{enumitem}
\usepackage{latexsym}
\usepackage{amsfonts}   
\usepackage{amssymb}
\usepackage{array}      
\usepackage{epsfig}
\usepackage[dvipsnames]{xcolor}
\usepackage{color}
\usepackage[colorlinks=true,linkcolor=blue,urlcolor=blue,citecolor=blue,pdfusetitle]{hyperref}
\usepackage{hyperref}
\usepackage[capitalize]{cleveref}
\usepackage{physics}
\usepackage{bbold}

\begin{document}

\title{Exponentially accelerated approach to stationarity in
Markovian open quantum systems through the Mpemba effect}
       
\author{Federico Carollo}
\affiliation{Institut f\"ur Theoretische Physik, Universit\"at T\"ubingen, Auf der Morgenstelle 14, 72076 T\"ubingen, Germany}

\author{Antonio Lasanta}
\affiliation{Departamento de \'Algebra, Facultad de Educaci\'on, Econom\'ia y Tecnolog\'ia de Ceuta, Universidad de Granada, E-51001 Ceuta, Spain}
\affiliation{G. Mill\'an Institute for Fluid Dynamics, Nanoscience and Industrial Mathematics,
Universidad Carlos III de Madrid, 28911 Legan\'es, Spain}

\author{Igor Lesanovsky}
\affiliation{Institut f\"ur Theoretische Physik, Universit\"at T\"ubingen, Auf der Morgenstelle 14, 72076 T\"ubingen, Germany}
\affiliation{School of Physics and Astronomy and Centre for the Mathematics and Theoretical Physics of Quantum Non-Equilibrium Systems, The University of Nottingham, Nottingham, NG7 2RD, United Kingdom}

\begin{abstract}
Ergodicity-breaking and slow relaxation are intriguing aspects of nonequilibrium dynamics both in classical and in quantum settings. These phenomena are typically associated with phase transitions, e.g. the emergence of metastable regimes near a first-order transition or scaling dynamics in the vicinity of critical points. Despite being of fundamental interest the associated divergent time scales are a hindrance when trying to explore steady-state properties. Here we show that the relaxation dynamics of Markovian open quantum systems can be accelerated exponentially by devising an optimal unitary transformation that is applied to the quantum system immediately before the actual dynamics. This initial ``rotation" is engineered in such a way that the state of the quantum system becomes orthogonal to the slowest decaying dynamical mode. We illustrate our idea --- which is inspired by the so-called Mpemba effect, i.e., water freezing faster when initially heated up --- by showing how to achieve an exponential speed-up in the convergence to stationarity in Dicke models, and how to avoid metastable regimes in an all-to-all interacting spin system.  
\end{abstract}

\date{\today}
\maketitle

\begin{figure}
	\includegraphics[width=\columnwidth]{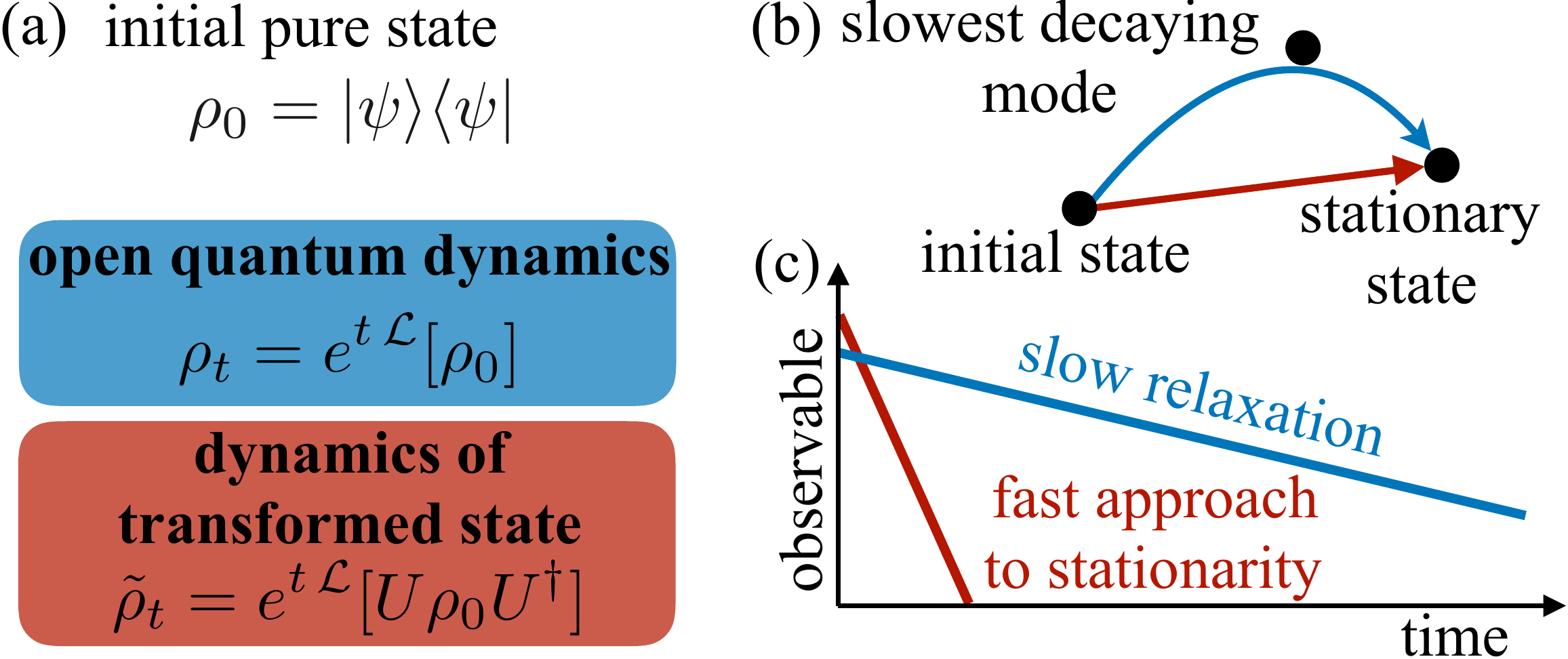}
	\caption{{\bf Mpemba effect in a Markovian open quantum system.} (a) We consider a quantum system, initially prepared in some pure state $\ket{\psi}$ and subject to a Markovian open quantum dynamics. Generically, the time scale for the approach to stationarity is contained in the dynamical generator $\mathcal{L}$ and is related to the slowest decaying excitation mode. Before the time evolution starts, we apply a unitary operation $U$ to the quantum state $\ket{\psi}$, which makes it ``orthogonal" to such slow mode. (b) After applying the unitary operation the system dynamics is not affected by long-lived excitation and approaches the stationary state in a ``more direct" way. (c) Sketch of the slow relaxation (blue line), contrasted with the accelerated one emerging after the applying the unitary (red line). The $y$-axis is in logarithmic scale.}
	\label{fig1}
\end{figure}

\noindent {\bf  Introduction.---} A strong focus of current research in many-body quantum physics is on understanding (nonequilibrium) phases of matter and transitions between them. Often associated with that are slow relaxation and divergent time-correlations \cite{polkovnikov2011,nandkishore2015,altman2015,turner2018,abanin2019,roy2020,pancotti2020}, which typically signal the onset of critical behavior \cite{hinrichsen2000,dagvadorj2015,carollo2019,gillman2019,jo2019,puel2021} or the appearance of metastable dynamical regimes \cite{bovier2002,macieszczak2016} near first order phase transitions. In certain instances, the concomitant very long relaxation time scales become impractical or even detrimental when a fast approach to stationarity is desired. This is certainly the case when one is interested in studying steady-state properties \cite{lin2013}, or, for instance, when the stationary state encodes the result of some computation \cite{verstraete2009,lewenstein2020}. In physical terms, the characteristic time needed for an open dissipative quantum system to approach stationarity is given by the lifetime $\tau$ of the slowest decaying excitation mode. A random initial pure state $\ket{\psi}$ [see Fig.~\ref{fig1}(a)] is typically out-of-equilibrium and excites all dynamical decaying modes, including the slowest one. As such, it will ultimately converge to stationarity in a time proportional to $\tau$. 

In this paper, we show that, if the slowest decaying mode is unique, one can always find an appropriate unitary operation which, once applied to the initial state $\ket{\psi}$, allows the open system to reach stationary behavior at a significantly faster pace. The idea, which is sketched in Fig.~\ref{fig1}(b), is that the unitary operation $U$ removes the excitation of the slowest decaying mode from the initial state $\ket{\psi}$, which achieves an exponential speed-up [see Fig.~\ref{fig1}(c)]. The basic mechanism underpinning our finding is reminiscent of the so-called Mpemba effect \cite{mpemba1969}, which refers to the phenomenon that a hotter liquid cools at a faster rate than a colder one. Often, this and related phenomena \cite{lu2017,lasanta2017,nava2019,baity-jesi2019,gal2020,kumar2020,Gijon2019,Klich2019,Torrente2019} indeed admit a transparent physical explanation \cite{lu2017,nava2019,gal2020,lu2017,Klich2019}: the state of the hotter system overlaps less with the slowest decaying modes of the cooling (dissipative) dynamics --- a hypothesis which has been confirmed experimentally in a trapped colloid particle  \cite{kumar2020}. In certain instances, however, such a clear separation of time scales may not occur. Still, an anomalous relaxation towards equilibrium can be investigated by monitoring the evolution of thermodynamic quantities representing internal degrees of freedom, such as kinetic and rotational energy, kurtosis or correlation length \cite{lasanta2017,baity-jesi2019,Gijon2019,Torrente2019}. 
Here, we explore the analogue of the Mpemba effect in Markovian open quantum systems. Using paradigmatic many-body systems of both theoretical and experimental interest we demonstrate the possibility of speeding-up the approach to stationarity and to avoid long pre-stationary metastable regimes. 

\noindent {\bf  Markovian open quantum dynamics.---} We first briefly discuss the fundamental elements of open quantum systems subject to a Markovian dynamics \cite{breuer2002,gardiner2004,lindblad1976,gorini1976,benatti2005,alicki2007,baumgartner2008,prosen2010,manzano2020}. The evolution of the density matrix $\rho_t$, describing the state of the quantum system, is governed by the quantum master equation $\dot{\rho}_t=\mathcal{L}[\rho_t]$ \cite{breuer2002,lindblad1976,gorini1976}, where $\mathcal{L}$ is the Lindblad map
\begin{equation}
\mathcal{L}[X]=-i[H,X]+\sum_{\mu=1}^{N_J}\left(L_\mu X L^\dagger_\mu-\frac{1}{2}\left\{L^\dagger_\mu L_\mu,X\right\}\right)\, .
\label{Lind}
\end{equation}
Here, $H=H^\dagger$ is the Hamiltonian of the system, and the $N_J$ jump operators $L_\mu$ describe the dissipative effects due to the presence of an environment. The Lindblad map $\mathcal{L}$ preserves the trace [$\Tr\left(\mathcal{L}[X]\right)=0$] and hermiticity [$\left(\mathcal{L}[X]\right)^\dagger=\mathcal{L}[X^\dagger]$, $\forall X$] and generates completely positive (physical) dynamics of the quantum state $\rho_t$.

The formal solution to the quantum master equation is given by $\rho_t=e^{t\mathcal{L}}[\rho_0]$,
where the exponential of the map $\mathcal{L}$ must be interpreted as the  power series.
Assuming the generator to be diagonalizable, one can find the right eigenmatrices, $r_k$, such that 
\begin{equation}
\mathcal{L}[r_k]=\lambda_k\, r_k\, .
\label{right-diag}
\end{equation}
The complex numbers $\lambda_k$ are the eigenvalues of the Lindblad map. Note that, due to the hermiticity-preservation of $\mathcal{L}$, if $\lambda_k$ is a complex eigenvalue, then $\lambda_k^*$ must also be an eigenvalue. For the same reason, one can also show that if $\lambda_k$ is real, then $r_k$ can be chosen to be Hermitian. Associated with the map defined in Eq.~\eqref{Lind}, there is a dual map, also called the adjoint Lindblad map, which implements the evolution of observables:
$$
\mathcal{L}^{+}[O]=i[H,O]+\sum_{\mu=1}^{N_J} \left(L^\dagger_\mu O L_\mu-\frac{1}{2}\left\{O,L^\dagger_\mu L_\mu\right\}\right).
$$
This dual map, $\mathcal{L}^+$, is diagonalized by the left eigenmatrices $\ell_k$, 
\begin{equation}
\mathcal{L}^+[\ell_k]=\lambda_k\, \ell_k\, .
\label{left-diag}
\end{equation}
The matrices $\ell_k$ are in principle different from the matrices $r_k$ in Eq.~\eqref{right-diag}. However, $\ell_k$ and $r_k$ still form a basis for the space of matrices and can always be defined with the property $\Tr\left(\ell_k r_h\right)=\delta_{kh}$. 

Since the dynamics generated by $\mathcal{L}$ is completely positive, the eigenvalues of the Lindblad map all have a non-positive real part, ${\rm Re}\left(\lambda_k\right)\le0$. Furthermore, trace preservation enforces that at least one eigenvalue is zero, $\lambda_1=0$. If such an eigenvalue is non-degenerate --- we will work under this assumption --- the (asymptotic) stationary state of the open quantum system,
\begin{equation}
\rho_{\rm ss}=\lim_{t\to\infty}\rho_t\, ,
\label{rho_ss}
\end{equation}
is unique and given by the right eigenmatrix $r_1$. Since the left eigenmatrix associated with $\lambda_1$ is the identity, $\ell_1={\bf 1}$, one has $\Tr \left(r_1\right)=1$. Finally, the matrix $r_1$ is guaranteed to be positive due to complete positivity of $e^{t\mathcal{L}}$.

The spectral decomposition of $\mathcal{L}$ allows us to write the dynamics of any initial density matrix as 
\begin{equation}
e^{t\mathcal{L}}\left[\rho_0\right]=r_1+\sum_{k=2}^{d^2}e^{t\lambda_k}{\rm Tr}\left(\ell_k\, \rho_0\right)r_k\, ,
\label{dyn}
\end{equation}
where $d$ is the dimension of the Hilbert space of the system. This decomposition shows that the matrices $r_k$ are nothing but the excitation modes of the system, each one characterized by its decay rate $|{\rm Re}(\lambda_k)|$. For long times, the relevant terms are those related to the $\lambda_k$ with the smallest real part in modulus. We order the eigenvalues $\lambda_k$ in such a way that $|{\rm Re}\left(\lambda_2\right)|\le |{\rm Re}\left(\lambda_3\right)|\le \dots \le |{\rm Re}\left(\lambda_m\right)|$ and we further assume that the eigenvalue $\lambda_2$ is real and unique. In this case, the time scale for relaxation is given by  
\begin{equation}
\tau=\frac{1}{|\lambda_2|}\, , 
\label{time-scale}
\end{equation}
and $r_2$ is in fact the slowest decaying excitation mode of the Markovian open quantum dynamics. 
 
\noindent {\bf Mpemba effect.---} A generic initial state will overlap with all decaying modes of a Lindblad dynamics, and thus, in particular, also with the slowest one. As such, the approach to the stationary state will take place in a time which is of the order of the relaxation time \eqref{time-scale}. However, looking at Eq.~\eqref{dyn}, one sees that this time scale becomes completely irrelevant for the dynamics if $\Tr\left(\ell_2 \rho_0\right)=0$. In such a case,  the state would relax at a faster rate with time-scale $1/|{\rm Re}(\lambda_3)|$, which implies an exponential speed up of the convergence to stationarity. In what follows, we show how such acceleration may always be achieved when starting from an initial pure state, $\rho_0=\ket{\psi}\bra{\psi}$, by performing a unitary rotation to the state before the actual time-evolution takes place. This is in spirit similar to the Mpemba effect, where an initial thermal state is first heated up before the cooling dynamics is started.

Given an initial pure state $\rho_{0}=\ket{\psi}\bra{\psi}$, there always exists a unitary $U$ --- which depends on the state --- such that
\begin{equation}
\Tr\left(\ell_2 \, U\rho_0 U^\dagger \right)=0\, ,
\label{goal}
\end{equation}
if the slowest decaying mode is unique. This can be shown as follows. First of all, we notice that the matrix $\ell_2$ must be Hermitian since we have assumed that $\lambda_2$ is real and nondegenerate. As such we can write it in its spectral form 
$$
\ell_2=\sum_{k=1}^d\alpha_k\ket{\varphi_k}\bra{\varphi_k}\, ,
$$
where $\bra{\varphi_k}\ket{\varphi_h}=\delta_{kh}$. We then note that, since ${\rm Tr}\left(\ell_2\, r_1\right)=0$ and $r_1$ is positive, the set of eigenvalues $\alpha_k$ must contain at least two eigenvalues with opposite sign or one equal to zero. Introducing an auxiliary orthonormal basis $\{\ket{\psi_k}\}_{k=1}^d$ for which $\ket{\psi}=\ket{\psi_1}$ (i.e. the initial state is a basis state) and using the spectral decomposition we find for the left hand side of Eq. (\ref{goal}):
$$
\Tr\left(\ell_2 \, U\rho_0 U^\dagger \right)=\sum_{k=1}^d\alpha_k\bra{\psi_1}U^\dagger\ket{\varphi_k}\bra{\varphi_k}U\ket{\psi_1}\, .
$$
To simplify the construction of the unitary we divide it into two parts, $U=U_2\, U_1$. The first unitary is chosen such that it maps the auxilliary basis $\ket{\psi_k}$ onto the basis $\ket{\varphi_k}$, which is simply achieved by  $U_1=\sum_k\ket{\varphi_k}\bra{\psi_k}$, yielding
$$
\Tr\left(\ell_2 \, U\rho_0 U^\dagger \right)=\sum_k\alpha_k\bra{\varphi_1}U_2^\dagger\ket{\varphi_k}\bra{\varphi_k}U_2\ket{\varphi_1}\, .
$$
In the next step we construct $U_2$ such that the right hand side of this expression becomes zero. Recalling that $\alpha_k$ are real numbers, two cases need to be considered: In case one of the $\alpha_k$ is zero, it is sufficient that $U_2$ performs a permutation of the basis $\{\ket{\varphi_k}\}$, mapping $\ket{\varphi_1}$ onto the eigenstate $\ket{\varphi_h}$ for which $\alpha_h=0$. 

In the non-trivial case, in which $\ell_2$ does not have a zero eigenvalue, we can make a construction based on the following observation: the eigenvalue $\alpha_1$ is a real number and can be either positive or negative. Since $\ell_2$ cannot be a positive (or negative) eigenmatrix there must be an eigenvalue $\alpha_n$ such that ${\rm sign}(\alpha_n)=-{\rm sign}(\alpha_1)$. We then construct the Hermitean operator $F=\ket{\varphi_1}\bra{\varphi_n}+\ket{\varphi_n}\bra{\varphi_1}$, 
which we use to define the unitary
\begin{equation}
\begin{split}
U(s):=e^{-is\, F}={\bf 1}+\left(\cos(s)-1\right) F^2 -i\sin(s)F
\end{split}
\end{equation}
where $F^2=\ket{\varphi_1}\bra{\varphi_1}+\ket{\varphi_n}\bra{\varphi_n}$. Using this unitary operator we find that 
\begin{equation}
\begin{split}
\Tr\left(\ell_2 \, U(s)U_1\rho_0 U_1^\dagger U^\dagger(s) \right)=\alpha_1\cos^2(s)+\alpha_n\sin^2(s)\, .
\end{split}
\label{fin-part}
\end{equation}
The above quantity has the same sign as $\alpha_1$ for $s=0$ but, on the other hand, it has the same sign as $\alpha_n$ for $s=\pi/2$. In particular, it vanishes for $\bar{s}=\arctan \left(\sqrt{\left|\alpha_1/\alpha_n\right|}\right)$, so that if we take the unitary $U=U(\bar{s})U_1$, Eq.~\eqref{goal} is satisfied. This implies that the initial state is rotated into a state which is orthogonal to the slowest decaying mode and will thus relax, in general, with the time scale $1/|{\rm Re}(\lambda_3)|$. In particular, this means that the approach to stationarity has been exponentially accelerated by a factor $|{\rm Re}(\lambda_3)|-|{\rm Re}(\lambda_2)|$.

\noindent{\bf Application to the dissipative Dicke model.---} As a first application of our result, we consider the single-mode open quantum Dicke model \cite{kirton2019,roses2020}, which is paradigmatic for the understanding of matter-light interactions and variants of which have been realised in a number of experiments \cite{dimer2007,nagy2010,baumann2010,ritsch2013,klinder2015}. It consists of an ensemble of two-level quantum systems, each of which is described by the spin operators $s_\alpha^{(k)}=\frac{1}{2}\sigma_{\alpha}^{(k)}$, with  $\sigma_\alpha$ being the Pauli matrix $\alpha$. The superscript $k$ indicates the spin to which the operator belongs. These spin variables are coupled to a bosonic mode, described by annihilation and creation operators, $a,a^\dagger$. 

In the Markovian regime, the open quantum dynamics of the Dicke model is described by a generator of the form in Eq.~\eqref{Lind}, with Hamiltonian \cite{kirton2019}
$$
H=\Omega S_z +\omega a^\dagger a\, +\frac{g}{\sqrt{N}}\left(a+a^\dagger \right)S_x\, ,
$$
and a single jump operator ($N_J=1$), $L_1=\sqrt{\kappa}a$. This latter contribution accounts for dissipative losses of excitations for the bosonic mode. While our method can also be applied to the above model, in order to simplify the numerics we make an assumption. We consider the adiabatic elimination of the bosonic mode.  By performing such an approximation (see Supplemental Material), the model is described solely in terms of spin degrees of freedom. The dynamics is governed by a generator of the form \eqref{Lind}, with
\begin{equation}
\tilde{H}=\Omega S_z-\frac{4\omega g^2}{4\omega^2 +\kappa^2}\frac{S_x^2}{N}\, ,\quad \tilde{L}_1=\frac{2|g|\sqrt{\kappa}}{\sqrt{4\omega^2+\kappa^2}}\frac{S_x}{\sqrt{N}}\, .
\label{spin-only-Hamiltonian}
\end{equation}
The above dynamics conserves the total angular momentum $S^2=S_x^2+S^2_y+S^2_z$. In the following we consider the largest symmetry sector, for which $S^2=N(N+1)/4$. This subspace is formed by the $2m+1$ eigenstates of the $S_z$ operator, $S_z\ket{m}=m\ket{m}$ with $m=-N/2,-N/2+1, \dots N/2$. 
\begin{figure}
	\includegraphics[width=\columnwidth]{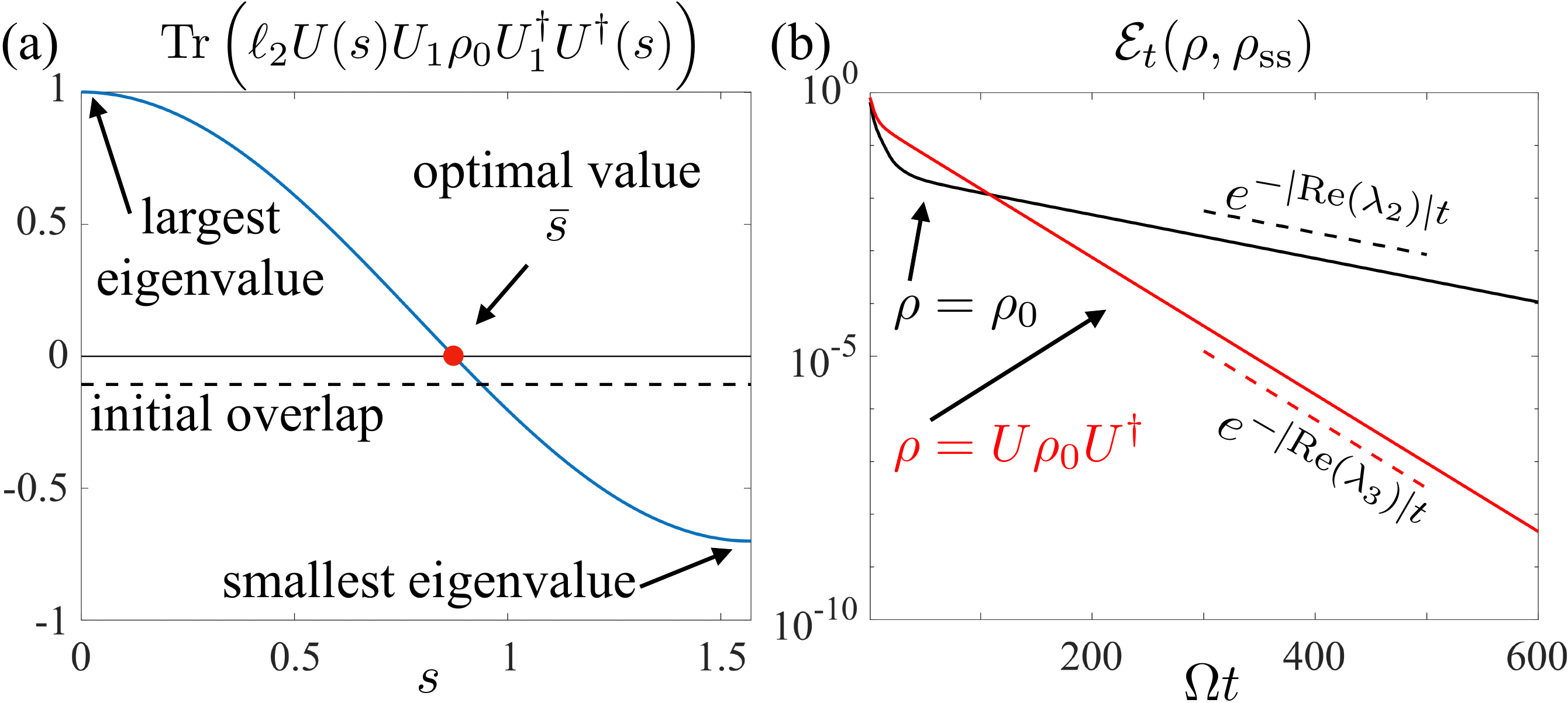}
	\caption{{\bf Dissipative Dicke model.} (a) Dependence of the overlap of a rotated initial random state on the rotation parameter $s$. According to Eq.~\eqref{fin-part}, this overlap can interpolate between the eigenvalue $\alpha_1$ of $\ell_2$ --- which we take here to be the largest in modulus --- and the eigenvalue $\alpha_n$, which is the largest one with sign opposite to $\alpha_1$. There always exists an optimal value $\bar{s}$ for which the overlap can be tuned to zero. The dashed line shows the overlap, $\bra{\psi}\ell_2\ket{\psi}$, of the initial random state with the decaying mode $r_2$. (b) Distance between the time-evolved state and the stationary state $\rho_{\rm ss}$. We compare the case of an initial random state (black line) with the time-evolution ensuing after the application of the rotation $U$ (see main text for discussion). While in the original case, the approach to stationarity is governed by the eigenvalue $\lambda_2$, the application of $U$ leads to an exponentially faster convergence to the steady-state with rate given by $|{\rm Re}(\lambda_3)|$. The parameters for this plot are $\omega=1$, $g=1$, $\kappa=1$ (all in units of $\Omega$) and $N=40$ spins.}
	\label{fig_Dicke}
\end{figure}
As initial state $\rho_0$ we take a random pure state of the form $\ket{\psi}\propto\sum_m (a_m+ib_m)\ket{m}$ with $a_m,b_m$ being uniformly distributed random numbers between $0$ and $1$. As shown in Fig.~\ref{fig_Dicke}(a), the overlap of a randomly selected state $\ket{\psi}$ with the matrix $\ell_2$ is generically finite. However, by tuning $U(s)$ we can find an appropriate transformation $U=U(\bar{s})U_1$ such that Eq.~\eqref{goal} is satisfied. For the rotated state, the overlap with $\ell_2$ is thus zero and we  have an approach to stationarity governed by the decay rate $|{\rm Re}(\lambda_3)|$. This is clearly shown in Fig.~\ref{fig_Dicke}(b), where we plot the Hilbert-Schmidt distance
\begin{equation}
\mathcal{E}_t(\rho,\rho_{\rm ss})=\left[\Tr\left(e^{t\, \mathcal{L}}\left[\rho\right]-\rho_{\rm ss}\right)^2\right]^{1/2}\, ,
\label{HS_distance}
\end{equation} 
between the stationary state $\rho_{\rm ss}$ [cf.~Eq.~\eqref{rho_ss}] and the time-dependent state starting from the state $\rho_0$ as well as from $U\rho_0 U^\dagger$, respectively. 
 
\begin{figure}
	\includegraphics[width=\columnwidth]{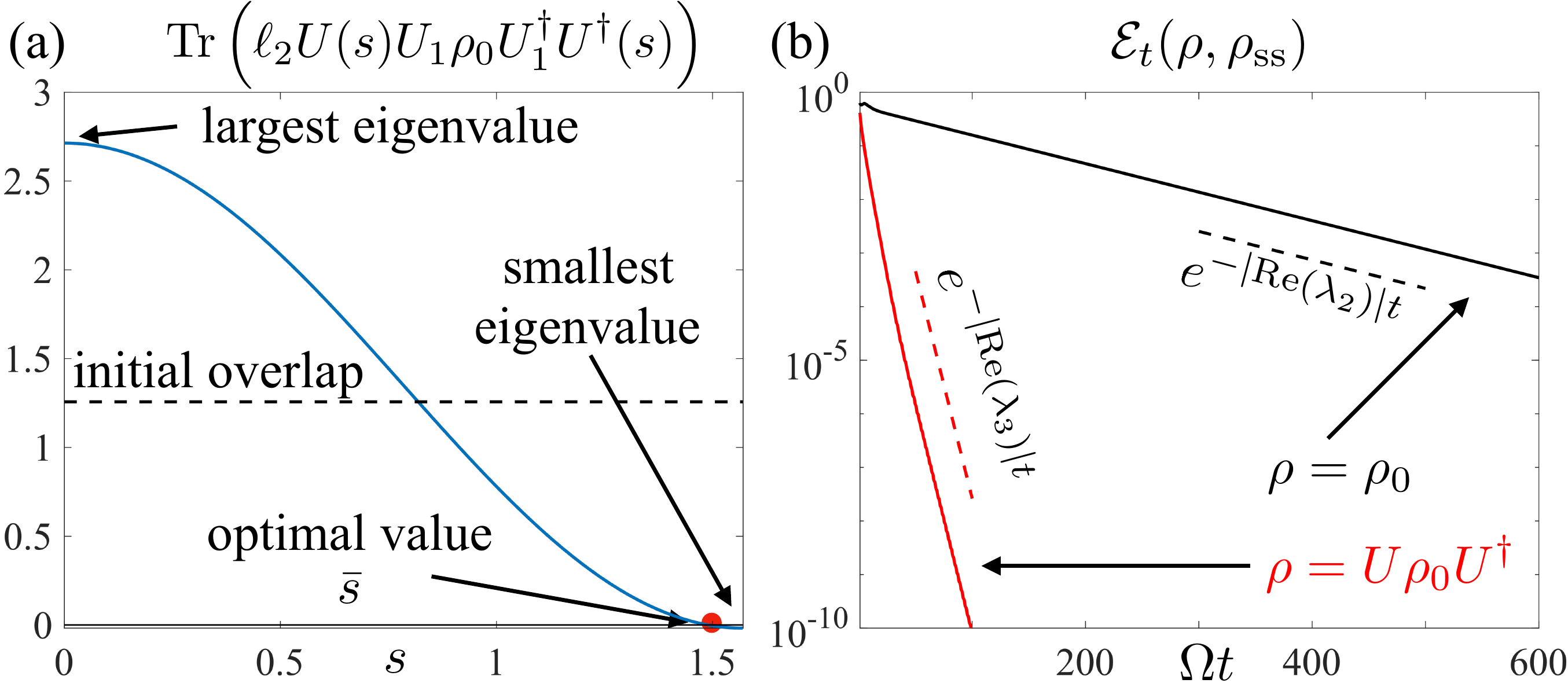}
	\caption{{\bf All-to-all interacting spin model.} (a) Dependence of the overlap of a rotated initial random state on the parameter $s$.  For $s=0$ this coincides with the largest eigenvalue of $\ell_2$, while for $s=\pi/2$ it coincides with the smallest. The latter eigenvalue is negative and very close to zero in modulus. Also in this case  there exists an optimal value $\bar{s}$ for which the overlap can be tuned to zero. The dashed line shows the overlap $\bra{\psi}\ell_2\ket{\psi}$. (b) Distance between the time-evolved state and the stationary state $\rho_{\rm ss}$ of the interacting spin system. We compare the case of the initial random state $\rho=\rho_0$ with the one obtained after the rotation $\rho=U\rho_0U^\dagger$. We see how in this case, the transformation $U$ prevents the system from entering a metastable regime which would slow down the approach to stationarity. The parameters for this plot are $\Delta=-1$, $V=3$, $\kappa=1$ (all in units of $\Omega$) and $N=40$ spins.}
	\label{fig_Rydberg}
\end{figure}

\noindent{\bf Application to an all-to-all interacting spin model.---} As a second application, we consider a spin model with resemblance to laser-driven interacting ensembles of Rydberg atoms \cite{bloch2008,saffman2010,lee2012,kshetrimayum2017,letscher2017}. This model allows us to demonstrate how the Mpemba effect may be used to avoid long-lasting metastable regimes \cite{macieszczak2016,rose2016,letscher2017}. From a theoretical perspective, we model the $N$ atoms as two-level systems, exactly through the spin degrees of freedom introduced previously. The state with spin pointing up in the $z$ direction corresponds to the atom being in the excited (Rydberg) state, while the one pointing down refers to the ground state of the atom. We consider a Markovian dynamics such as the one in Eq.~\eqref{Lind} with
$$
H=\Omega S_x-\Delta S_z+\frac{V}{N}S_z^2\, , \quad L_1=\sqrt{\kappa}S_-\, ,
$$
where $S_-=S_x-iS_y$ is a spin ladder operator. For this model, $\Omega$ is the Rabi frequency, $\Delta$ is the laser detuning with respect to the atomic transition frequency $\omega_{\rm at}$ while $V$ parametrises here the strength of the all-to-all interactions. 

For certain parameters regimes, e.g.~the one considered in Fig.~\ref{fig_Rydberg}, the model features a so-called metastable regime, which emerges since $|{\rm Re}(\lambda_2)|\ll |{\rm Re}(\lambda_3)|$. This means that, over a long time window during which all other decaying excitation modes have already relaxed, the mode related to $|{\rm Re}(\lambda_2)|$ is still relevant and keeps the system away from stationarity. In such a scenario, the accelerated relaxation achieved by applying the transformation $U$ is even more striking since the exponential gain is by a factor $e^{t\left(|{\rm Re}(\lambda_3)|-|{\rm Re}(\lambda_2)|\right)}$. This is can be appreciated from the curves displayed in Fig.~\eqref{fig_Rydberg}.

\noindent{\bf Discussion.---} We have presented a general method to control the time scale for the approach to stationarity in Markovian open quantum systems, which can be considered a quantum variant of the so-called Mpemba effect. Our results show how a dramatically accelerated approach to stationarity can be achieved by applying a suitable unitary transformation to the initial state, which removes its overlap with the slowest decaying mode. We note that the unitary operation $U$ introduced in this work is ``optimal" in the sense that it completely de-populates the slowest decaying mode. However, as discussed for instance in Ref.~\cite{nava2019}, in order to observe a Mpemba effect it would be sufficient to engineer a rotation which simply diminishes the excitation of such a slow mode. As shown in Fig.~\ref{fig_Dicke}-\ref{fig_Rydberg}, this is also achieved by ``non-optimal" values of the parameter $s$ for which the transformed overlap is smaller, in modulus, than the initial one. Considering specific many-body quantum models, it would be interesting to explore the possibility to reduce the population of the slowest decaying mode by means of a less involved unitary, for instance by implementing local and independent rotation of the different system constituents. This would not lead to an ``optimal" speed-up but would facilitate the implementation and observation of the Mpemba effect in actual experiments with open quantum many-body systems.

\acknowledgments
\noindent{\bf Acknowledgements.---} F.C. and I.L. acknowledge support from the “Wissenschaftler-R\"uckkehrprogramm GSO/CZS” of the Carl-Zeiss-Stiftung and the German Scholars Organization e.V., as well as through the Deutsche Forschungsgemeinsschaft (DFG, German Research Foundation) under Project No. 435696605. A.L. acknowledges support from the Spanish Ministerio de Ciencia, Innovaci\'on y Universidades and the Agencia Estatal de Investigaci\'on  through the grant No. MTM2017-84446-C2-2-R.

\bibliographystyle{apsrev4-1}
\bibliography{Mpemba}

\setcounter{equation}{0}
\setcounter{figure}{0}
\setcounter{table}{0}
\setcounter{page}{1}
\makeatletter
\renewcommand{\theequation}{S\arabic{equation}}
\renewcommand{\thefigure}{S\arabic{figure}}
\renewcommand{\bibnumfmt}[1]{[S#1]}
\renewcommand{\citenumfont}[1]{S#1}
\makeatletter
\renewcommand{\theequation}{S\arabic{equation}}
\renewcommand{\thefigure}{S\arabic{figure}}
\renewcommand{\bibnumfmt}[1]{[S#1]}
\renewcommand{\citenumfont}[1]{S#1}

\newpage
\onecolumngrid

\section{SUPPLEMENTAL MATERIAL}
\label{App:adiabatic-elimination}
We show here how to obtain the spin-only description of the Dicke model by performing an adiabatic elimination of the bosonic mode. The starting point is to consider the Heisenberg equation of motion for the annihilation (or creation) operator. This equation reads as 
\begin{equation}
\frac{d}{dt}a=\mathcal{L}^+[a]=\left(-i\omega -\frac{\kappa}{2}\right)a-i\frac{g}{\sqrt{N}}S_x\, .
\label{Heisenberg-a}
\end{equation}
The actual approximation consists in setting the above derivative to zero. This produces a relation between the bosonic operator $a$ and the collective spin operator $S_x$. The idea behind is that the dynamics, and in particular the decay time-scale, of the bosonic mode is much faster than the dynamics on the spin degrees of freedom. In this regime, the bosonic mode can be ``slaved" to the spin operator $S_x$. Setting Eq.~\eqref{Heisenberg-a} to zero, one indeed finds the relation
\begin{equation}
a=-\frac{g\left(4\omega +2i\kappa \right)}{\sqrt{N}\left(4\omega^2 +\kappa^2\right)}S_x\, .
\label{adiabatic-elimination}
\end{equation}
As a consequence, we can write 
$$
(a+a^\dagger)=-\frac{8g\omega}{\sqrt{N}\left(4\omega^2+\kappa^2\right)}S_x\, ,
$$
as well as 
$$
\omega a^\dagger a=\frac{4\omega g^2}{4\omega^2+\kappa^2}\frac{S_x^2}{N}\, .
$$
Substituting these relations in the Lindblad generator of the Dicke model, we find the spin-only dynamical description given in the main text.

\end{document}